\documentclass[preprint,12pt]{elsarticle}

\usepackage{amssymb}

\newcounter{bla}

\journal{Computer Physics Communications}

\begin{document}

\begin{frontmatter}

\title{Drawing Feynman diagrams with GLE}

\author{Andrey~Grozin}
\ead{A.G.Grozin@inp.nsk.su}
\affiliation{organization={Budker Institute of Nuclear Physics},
addressline={Lavrentiev st. 10},
postcode={630090},
city={Novosibirsk},
country={Russia}}

\begin{abstract}
A package for drawing publication-quality Feynman diagrams written in GLE is described.
\end{abstract}


\end{frontmatter}



{\bf PROGRAM SUMMARY}

\begin{small}
\noindent
{\em Program Title:} feyn.gle\\
{\em CPC Library link to program files:} (to be added by Technical Editor)\\
{\em Code Ocean capsule:} (to be added by Technical Editor)\\
{\em Licensing provisions:} BSD 3-clause\\
{\em Programming language:} GLE~[1]\\
{\em Supplementary material:}                                 \\
{\em Nature of problem:}\\
Feynman diagrams are widely used in publications on quantum field theory,
elementary particle physics, and some other areas of physics.
They are needed in articles, books, lecture courses, conference presentations, web pages, etc.
General-purpose graphics programs are poorly suited
for production of publication-quality diagrams.\\
{\em Solution method:}\\
A package of subroutines in the GLE programming language
for drawing Feynman diagrams is provided.
GLE~[1] is a free cross-platform graphics program
which transforms scripts written in the GLE language
to graphics files of various kinds (eps, ps, pdf, svg, jpg, png).
These files can be included in \LaTeX{} documents,
documents in various office suites (like Word, PowerPoint, etc.),
or web pages.
\\

\end{small}

\section{Introduction}
\label{S:Intro}

There are many programs for drawing Feynman diagrams.
They can be classified into GUI programs and script-based ones
(where the author has to write code in some language to produce a diagram).
Some programs for generating and manipulating Feynman diagrams can also visualize them;
we don't discuss them here, because they don't let the author to produce an arbitrary diagram.

\begin{sloppypar}
Several \LaTeX{} packages allow the author to write code producing diagrams
directly in a \LaTeX{} file
(we don't discuss packages incompatible with modern versions of \LaTeX).
One of them is \verb|Axodraw|~\cite{Vermaseren:1994je,Collins:2016aya}
(\verb|https://ctan.org/pkg/axodraw|, \verb|https://ctan.org/pkg/axodraw2|).
It inserts PostScript code for drawing a diagram into the dvi file
using \TeX{} \verb|\special|,
and therefore requires one to use \verb|latex| and \verb|dvips| to produce a PostScript file
containing diagrams (which can later be transformed to pdf by \verb|ps2pdf|).
The version 2 allows one to produce pdf directly using \verb|pdflatex| or \verb|lualatex|,
but requires one to use a separate program \verb|axohelp| in the process.
The package \verb|pst-feyn| (\verb|https://ctan.org/pkg/pst-feyn|)
is based on the old version of \verb|Axodraw|, but uses \verb|PSTricks|.
\end{sloppypar}

The package \verb|feynMF|~\cite{Ohl:1995kr} (\verb|https://ctan.org/pkg/feynmf|)
uses \textsf{METAFONT} to generate a bitmap font containing needed diagrams.
Its variant \verb|feynMP| uses \textsf{METAPOST} instead to produce eps files
of the diagrams.
This package does not require the author to specify coordinates of all vertices in a diagram,
it uses an algorithm (which can be fine-tuned) to find these coordinates.

\begin{sloppypar}
The package \verb|TikZ-Feynman|~\cite{Ellis:2016jkw} (\verb|https://ctan.org/pkg/tikz-feynman|)
uses \verb|PGF/TikZ|.
It can use several algorithms to find optimal vertex coordinates,
and needs \verb|lualatex| to do so.
Vertex coordinates can be also specified by hand.
A variant of this package \verb|TikZ-Feynhand| (\verb|https://ctan.org/pkg/tikz-feynhand|)
allows only setting vertices by hand, but can work with the ordinary \verb|pdflatex|.
\end{sloppypar}

\begin{sloppypar}
There are also script-based packages based on languages other than \LaTeX.
\verb|asymptote| (\verb|https://asymptote.sourceforge.io/|) is distributed with such a package,
but its functionality is rather limited.
\verb|PyFeyn| (\verb|https://pyfeyn.hepforge.org/|)
is written in python and based on \verb|PyX|;
unfortunately, it does not work with the current version of \verb|PyX|,
and can no longer be used.
\verb|FeynDiagram| (\verb|http://www.feyndiagram.com/|)
is a library of \verb|C++| classes for producing PostScript files with diagrams.
\end{sloppypar}

Let's also mention a few GUI programs.
\verb|Xfey|~\cite{Laina:1998ek} is written in \verb|C++|
and uses \verb|Motif| and \verb|Xmt| library;
nowadays it is not easy to compile it.
Its functionality is rather limited.
\verb|JaxoDraw|~\cite{Binosi:2003yf,Binosi:2008ig} is written in \verb|Java|,
it can be used as a GUI front-end for \verb|Axodraw| or separately.
\verb|FeynGame|~\cite{Harlander:2020cyh} is also written in \verb|Java|.
Finally, \verb|https://feynman.aivazis.com/| is a web page allowing one to draw
Feynman diagrams in a browser.

The package \verb|feyn.gle| was written by me in 1993.
Its version 1.0 was distributed with GLE for many years.
The package was used for many years for drawing Feynman diagrams
for journal articles, books, conference presentations, and lecture courses.
However, there was no manual documenting it, and therefore it was not widely used.
Recently I substantially updated it to the version 1.1.3.
This version is described in the current article.
Example programs from it will not work with the version 1.0 currently available in GLE.
The new version will be included in a future version of GLE.
Before this, you can download it from \verb|https://www.inp.nsk.su/~grozin/feyn_gle/|
and put to some directory used by \verb|gle| for searching \verb|include| files.
This means a current directory where you use \verb|gle|;
a directory defined in the environment variable \verb|GLE_USRLIB|;
or the directory \verb|gleinc| in the GLE installation.

Features of this package are somewhat similar to \verb|Axodraw|.
It cannot choose vertices' coordinates according to some algorithm,
as \verb|feynMF| and \verb|TikZ-Feynman|.
All coordinated are provided by the user.
However, the author usually has a rather precise idea how his diagrams should look.
Tuning an algorithm to produce something in agreement with such expectation
can become more difficult than specifying all coordinates.
GLE is a full-fledged programming language,
and some coordinates can be calculated according to non-trivial formulas
from analytic geometry (in \LaTeX{} this would be more difficult).

This package can produce graphics files with diagrams
of all types available in GLE.
These files can be included in a \LaTeX{} document (inside figures or equations)
using the standard \LaTeX{} packages \verb|graphics| or \verb|graphicx|.
In many cases it is better to include graphics files with diagrams
inside \verb|picture| environments using \verb|\put| to place them at desired positions.
Other \verb|\put| commands can add text fragments and formulas (e.g., particle names and momenta)
inside the diagrams.
In this way, the text and formula fonts in figures
will be the same as used in the main text.

\verb|feyn.gle| is easy to use and flexible.
For example, the following program produces the diagram shown below:\\
\begin{minipage}[t]{0.48\textwidth}
\small
\begin{verbatim}
x = 1
y = 3/4*x
x1 = 1.6
d = 0.1

size 2*(x1+d) 2*(y+d)
include feyn.gle

begin translate x1+d y+d
    amove -x1 0
    @Photon -x 0
    @Vert
    @Fermion2 0 y x 0
\end{verbatim}
\end{minipage}
\begin{minipage}[t]{0.48\textwidth}
\small
\begin{verbatim}
    @Arrow 1/4
    @Arrow 3/4
    @Vert
    @Fermion2 0 -y -x 0
    @Arrow 1/4
    @Arrow 3/4
    amove x 0
    @Photon x1 0
    amove 0 -y
    @Vert
    @Photon 0 y
    @Vert
end translate
\end{verbatim}
\end{minipage}\\
\includegraphics{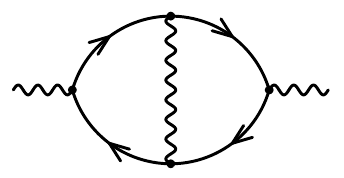}\\
If you want to tweak its shape and size,
just edit the parameters in the first few lines and re-run gle.

If your diagram is in the file \verb|dia.gle| then running
\begin{verbatim}
gle dia.gle
\end{verbatim}
produces the encapsulated PostScript file \verb|dia.eps|.
It can be included in a \LaTeX{} file
if you plan to compile it with the command \verb|latex|.
Running
\begin{verbatim}
gle -d pdf dia.gle
\end{verbatim}
produces the pdf file \verb|dia.pdf|.
It can be included in a \LaTeX{} file
if you plan to compile it with \verb|pdflatex| (or \verb|lualatex|).

GLE is case insensitive, so, you may write \verb|@Fermion| or \verb|@fermion| or any other version.
The package \verb|feyn.gle| contains a number of subroutines
and global variables which change their behavior in some ways;
all of them are described in this manual.
If there is a global parameter \verb|Foo|,
then usually the variable \verb|DefFoo| contains its default value.
So, you can easily do \verb|Foo = DefFoo|
after any changes.
There are a few exceptions:
if the default value of \verb|Foo| is 0 or 1 (dimensionless, not a length),
or the color \verb|"black"| or \verb|"white"|,
then there is no \verb|DefFoo|
(it is easy enough to write \verb|Foo = 0| or similar).
It also uses a number of global variables for its internal working.
Names of all such variables begin with \verb|f|.
So, you should not use any variables beginning with \verb|f| or \verb|F|
in a program which includes this package.

It is convenient to introduce parameters and to assign values to them
at the top of the file.
All coordinates and other quantities should be expressed via these parameters.
Then it is easy to tune your drawing by changing values of these parameters ---
no need to edit something inside the file, perhaps in many places.
It is also convenient to put each part of the drawing which can be moved
(probably, each separate diagram)
inside a\\
\texttt{begin translate \boldmath $x$ $y$}\\
\verb|   |$\cdots$\\
\texttt{end translate}\\
block.
The picture created by the code inside this block is translated by $(x,y)$.
It can be easily moved to another place by changing $x$ and $y$,
without editing the code inside.
Coordinates of some points in a diagram may be results of some calculations.
If there are bugs in such formulas the resulting diagram may look weird.
In such cases you can use \verb|print| for printing results of such calculations.
They will appear in the terminal from which you started \verb|gle|,
and you will be able to inspect them and see if they look plausible.

\section{Straight lines}
\label{S:Straight}

\texttt{\bf @Fermion \boldmath $x$ $y$}
draws a fermion line from the current point to $(x,y)$.
First you move the current point to $(x_0,y_0)$ by \texttt{amove \boldmath $x_0$ $y_0$};
then you draw a line to $(x,y)$,
after that $(x,y)$ becomes the current point.
After that you can draw another line, it will begin at the end of the previous line, etc.
You can produce a thick line using \verb|set lwidth|;
don't forget to reset it to the default value after that,
otherwise all subsequent drawings will be affected.\\
\begin{minipage}[t]{0.48\textwidth}
\small
\begin{verbatim}
begin translate 0.1 0.1
    amove 0 0
    @Fermion 2 0
end translate
\end{verbatim}
\end{minipage}
\begin{minipage}[t]{0.48\textwidth}
\small
\begin{verbatim}
begin translate 3.1 0.1
    amove 0 0
    set lwidth 2*DefLWidth
    @Fermion 2 0
    set lwidth DefLWidth
end translate
\end{verbatim}
\end{minipage}\\
\includegraphics{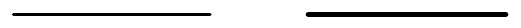}

\texttt{\bf @Vert}
draws a vertex at the current point.
You can change its radius.
A vertex is a circle (line) with the interior filled by the color \verb|FillC$|.
The line color can be changed by the GLE built-in command \verb|set color|;
the filling color --- by assignment to  \verb|FillC$|.
If these colors differ a vertex looks really strange
(though one may want it for an unusual effect).
It is better to change them in sync.
This is done by \verb|@SetColor|.\\
\begin{minipage}[t]{0.48\textwidth}
\small
\begin{verbatim}
begin translate 0.1 0.1
    amove 0 0
    @Vert
end translate

begin translate 1.1 0.1
    amove 0 0
    DotR = 2*DefDotR
    @SetColor "gray30"
    @Vert
    DotR = DefDotR
    @SetColor "black"
end translate
\end{verbatim}
\end{minipage}
\begin{minipage}[t]{0.48\textwidth}
\small
\begin{verbatim}
begin translate 2.1 0.1
    amove 0 0
    @Vert
    @Fermion 2 0
    @Vert
end translate
\end{verbatim}
\includegraphics{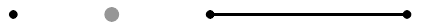}
\end{minipage}

\texttt{\bf @Arrow \boldmath $f$}
draws an arrow at the fraction $f$ of the line.
You can change its length and amplitude;
you can also displace it in the transverse direction
(positive means left), though this is rarely needed.\\
\begin{minipage}[t]{0.48\textwidth}
\small
\begin{verbatim}
begin translate 0.1 0.1
    amove 0 0
    @Fermion 2 0
    @Arrow 1/2
end translate

begin translate 3.1 0.1
    amove 0 0
    @Fermion 2 0
    ArrowL = 3/2*DefArrowL
    @Arrow 1/3
    ArrowL = DefArrowL
    ArrowA = 3/2*DefArrowA
    @Arrow 2/3
    ArrowA = DefArrowA
end translate
\end{verbatim}
\end{minipage}
\begin{minipage}[t]{0.48\textwidth}
\small
\begin{verbatim}
begin translate 6.1 0.1
    amove 0 0
    @Fermion 2 0
    ArrowD = 0.1
    @Arrow 1/2
    ArrowD = 0
end translate
\end{verbatim}
\end{minipage}\\
\includegraphics{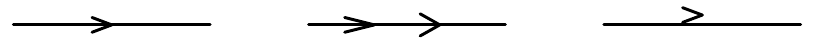}

\texttt{\bf @Mom \boldmath $f$}
draws an arrow for the momentum of a line at the fraction $f$ of its length.
It is displaced by \verb|MomD| in the transverse direction (positive means left)
and has length $2*$\verb|MomL|.
Options of \verb|Arrow| also influence it.\\
\begin{minipage}[t]{0.48\textwidth}
\small
\begin{verbatim}
begin translate 0.1 0.35
    amove 0 0
    @Fermion 2 0
    @Mom 1/2
end translate

begin translate 3.1 0.35
    amove 0 0
    @Fermion 2 0
    MomD = -DefMomD
    @Mom 1/2
    MomD = DefMomD
end translate
\end{verbatim}
\end{minipage}
\begin{minipage}[t]{0.48\textwidth}
\small
\begin{verbatim}
begin translate 6.1 0.35
    amove 0 0
    @Fermion 2 0
    MomL = 3/2*DefMomL
    @Mom 1/2
    MomL = DefMomL
end translate
\end{verbatim}
\end{minipage}\\
\includegraphics{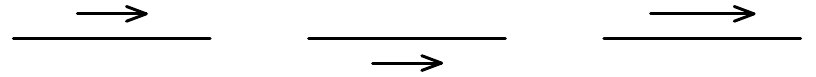}

Both \verb|@Arrow| and \verb|@Mom| produce arrows \emph{in the direction} of the line.
Sometimes one may want an arrow in the \emph{opposite} direction.
This can be achieved as\\
\begin{minipage}[b]{0.48\textwidth}
\small
\begin{verbatim}
begin translate 0.1 0.1
    amove 0 0
    @Fermion 2 0
    @Mom 1/2
    @Feyn 0 0
    @Arrow 1/2
end translate
\end{verbatim}
\end{minipage}
\begin{minipage}[b]{0.48\textwidth}
\includegraphics{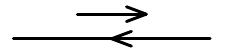}
\end{minipage}\\
\texttt{\bf @Feyn \boldmath $x$ $y$}
draws an \emph{invisible} line from the current point to $(x,y)$.

\texttt{\bf @Double \boldmath $x$ $y$}
draws a double line from the current point to $(x,y)$.
Such lines are used for HQET heavy quarks or Wilson lines.
The distance between lines is $2*$\verb|DoubleA|.
Note that $\mathtt{DefDoubleA} = \mathtt{DefDotR}$,
so that double lines blend with vertices nicely.
If you change one of them, it is a good idea to change the other one, too.\\
\begin{minipage}[t]{0.48\textwidth}
\small
\begin{verbatim}
begin translate 0.1 0.1
    amove 0 0
    @Vert
    @Double 2 0
    @Vert
end translate
\end{verbatim}
\end{minipage}
\begin{minipage}[t]{0.48\textwidth}
\small
\begin{verbatim}
begin translate 3.1 0.1
    amove 0 0
    DotR = 3/2*DefDotR
    DoubleA = 3/2*DefDoubleA
    @Vert
    @Double 2 0
    @Vert
    DotR = DefDotR
    DoubleA = DefDoubleA
end translate
\end{verbatim}
\end{minipage}\\
\includegraphics{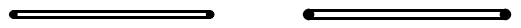}

There is a possibility to make the ends of a double line skew.
At the beginning of the double line
the beginning of its left line shifts by $\mathtt{DoubleB} * \mathtt{DoubleA}$ backward,
and the beginning of its right line shifts by the same amount forward.
At the end of the double line
the end of its left line shifts by $\mathtt{DoubleB} * \mathtt{DoubleA}$ forward,
and the end of its right line shifts by the same amount backward.
This is especially useful if you want to produce an angle made of double lines.
After the call to \verb|Double| both options return to their default value $0$ automatically.
So, you will have to set them again before the next call to \verb|Double|.\\
\begin{minipage}[b]{0.48\textwidth}
\small
\begin{verbatim}
begin translate 0.1 0.1
    amove 0 0
    DoubleB = -1
    DoubleE = 1
    @Double 1.5 1.5
    DoubleB = -1
    DoubleE = 1
    @Double 3 0
end translate
\end{verbatim}
\end{minipage}
\begin{minipage}[b]{0.48\textwidth}
\includegraphics{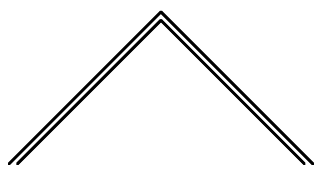}
\end{minipage}\\
Another way to make an angle formed by double lines look nice
is to put \verb|Vert| at the angle;
then there is no need to make ends skew.

\texttt{\bf @DMom \boldmath $f$}
draws an arrow similar to \texttt{\bf @Mom \boldmath $f$} but double.
Such arrows are sometimes used to show helicities of external lines.
Options \verb|DoubleA|, \verb|MomL|, \verb|MomD|, \verb|ArrowA|, \verb|ArrowL|
influence such arrows.\\
\begin{minipage}[b]{0.48\textwidth}
\small
\begin{verbatim}
begin translate 0.1 0.1
    amove 0 0
    @Fermion 2 0
    @Arrow 1/2
    @DMom 1/2
end translate
\end{verbatim}
\end{minipage}
\begin{minipage}[b]{0.48\textwidth}
\includegraphics{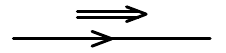}
\end{minipage}

\texttt{\bf @Dash \boldmath $x$ $y$}
draws a dashed line from the current point to $(x,y)$.
Such lines are used for Higgs.
The period is \verb|DashL|;
its fraction \verb|DashF| is occupied by the dash
(and hence the fraction $1-\mathtt{DashF}$ is a gap).\\
\begin{minipage}[t]{0.48\textwidth}
\small
\begin{verbatim}
begin translate 0.1 0.1
    amove 0 0
    @Dash 2 0
end translate

begin translate 3.1 0.1
    amove 0 0
    DashL = 1.5*DefDashL
    @Dash 2 0
end translate
\end{verbatim}
\end{minipage}
\begin{minipage}[t]{0.48\textwidth}
\small
\begin{verbatim}
begin translate 6.1 0.1
    amove 0 0
    DashF = 2/3
    @Dash 2 0
    DashL = DefDashL
    DashF = DefDashF
end translate
\end{verbatim}
\end{minipage}\\
\includegraphics{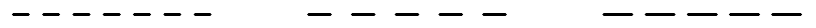}

If $\mathtt{DashB} > 0$,
it means that the beginning of the line is a dash of the length \verb|DashB| times the normal dash length.
If $\mathtt{DashB} < 0$,
it means that the beginning of the line is a gap of the length $|\mathtt{DashB}|$ times the normal gap length.
The variable \verb|DashE| controls the end of the line in the same way.
By default, both are 1.\\
\begin{minipage}[t]{0.48\textwidth}
\small
\begin{verbatim}
begin translate 0.1 0.6
    amove 0 0
    @Vert
    @Dash 2 0
    @Vert
end translate

begin translate 0.1 0.1
    amove 0 0
    @Vert
    DashB = -1
    @Dash 2 0
    DashB = 1
    @Vert
end translate
\end{verbatim}
\end{minipage}
\begin{minipage}[t]{0.48\textwidth}
\small
\begin{verbatim}
begin translate 3.1 0.6
    amove 0 0
    @Vert
    DashE = -1
    @Dash 2 0
    @Vert
end translate

begin translate 3.1 0.1
    amove 0 0
    @Vert
    DashB = -1
    @Dash 2 0
    @Vert
end translate
\end{verbatim}
\end{minipage}\\
\includegraphics{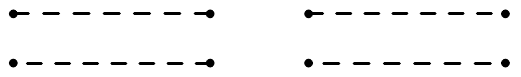}\\
\begin{minipage}[t]{0.48\textwidth}
\small
\begin{verbatim}
begin translate 0.1 0.6
    amove 0 0
    @Vert
    DashB = 0.5
    DashE = 0.5
    @Dash 2 0
    @Vert
end translate

begin translate 0.1 0.1
    amove 0 0
    @Vert
    DashB = -0.5
    @Dash 2 0
    @Vert
end translate
\end{verbatim}
\end{minipage}
\begin{minipage}[t]{0.48\textwidth}
\small
\begin{verbatim}
begin translate 3.1 0.6
    amove 0 0
    @Vert
    DashB = 0.5
    DashE = -0.5
    @Dash 2 0
    @Vert
end translate

begin translate 3.1 0.1
    amove 0 0
    @Vert
    DashB = -0.5
    @Dash 2 0
    DashB = 1
    DashE = 1
    @Vert
end translate
\end{verbatim}
\end{minipage}\\
\includegraphics{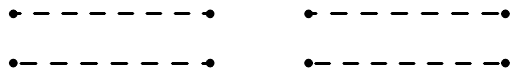}

The value of \verb|DashN| is added to the number of dashes calculated by the program.
If you set it to $+1$ (or $-1$) the parity is reversed.
After the call to \verb|Dash| this variable is automatically reset to its default value 0.\\
\begin{minipage}[t]{0.48\textwidth}
\small
\begin{verbatim}
begin translate 0.1 0.4
    amove 0 0
    @Dash 2 0
    amove 1 -0.3
    @Fermion 1 0.3
end translate
\end{verbatim}
\end{minipage}
\begin{minipage}[t]{0.48\textwidth}
\small
\begin{verbatim}
begin translate 3.1 0.4
    amove 0 0
    DashN = 1
    @Dash 2 0
    amove 1 -0.3
    @Fermion 1 0.3
end translate
\end{verbatim}
\end{minipage}\\
\includegraphics{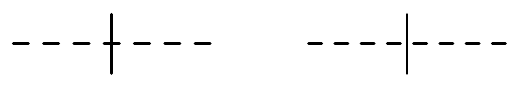}

\texttt{\bf @Dots \boldmath $x$ $y$}
draws a dotted line from the current point to $(x,y)$.
The period is \verb|DotsL|;
dots are controlled by \verb|DotR|,
their color can be changed in the same way as for \verb|Vert|.
Dots are \emph{not} drawn at the beginning of the line and at its end
(here the beginning and the end of each line are shown by cross markers).\\
\begin{minipage}[t]{0.48\textwidth}
\small
\begin{verbatim}
begin translate 0.1 0.1
    amove 0 0
    marker cross 0.5
    @Dots 2 0
    marker cross 0.5
end translate

begin translate 3.1 0.1
    amove 0 0
    marker cross 0.5
    DotsL = 2*DefDotsL
    @Dots 2 0
    DotsL = DefDotsL
    marker cross 0.5
end translate
\end{verbatim}
\end{minipage}
\begin{minipage}[t]{0.48\textwidth}
\small
\begin{verbatim}
begin translate 6.1 0.1
    amove 0 0
    marker cross 0.5
    DotR = 1.5*DefDotR
    @Dots 2 0
    DotR = DefDotR
    marker cross 0.5
end translate
\end{verbatim}
\end{minipage}\\
\includegraphics{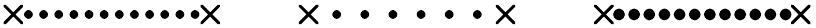}

\texttt{\bf @Zigzag \boldmath $x$ $y$}
draws a zigzag line from the current point to $(x,y)$.
Such lines are used for vector bosons ($W$, $Z$).
The ``half wave length'' is \verb|PhotonL|,
and the ``amplitude'' is \verb|PhotonA|.
If $\mathtt{PhotonA} > 0$ the zigzag deviates to the left from the straight line
at its beginning.\\
\begin{minipage}[t]{0.48\textwidth}
\small
\begin{verbatim}
begin translate 0.1 0.6
    amove 0 0
    @Zigzag 2 0
end translate

begin translate 0.1 0.1
    amove 0 0
    PhotonA = -DefPhotonA
    @Zigzag 2 0
    PhotonA = DefPhotonA
end translate
\end{verbatim}
\end{minipage}
\begin{minipage}[t]{0.48\textwidth}
\small
\begin{verbatim}
begin translate 3.1 0.6
    amove 0 0
    PhotonL = 1.5*DefPhotonL
    @Zigzag 2 0
    PhotonL = DefPhotonL
end translate

begin translate 3.1 0.1
    amove 0 0
    PhotonA = 1.5*DefPhotonA
    @Zigzag 2 0
    PhotonA = DefPhotonA
end translate
\end{verbatim}
\end{minipage}\\
\includegraphics{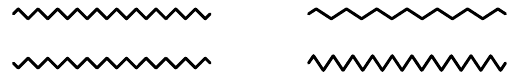}

\verb|Zigzag| ensures that the line contains an integer number of half-waves.
Their number can be either even
(integer number of waves,
the line leaves its beginning to the left and comes to its end from the right,
or wise versa)
or odd
(half-integer number of waves,
the line leaves its beginning to the left and comes to its end from the left,
or vice versa).
Sometimes one of these variants looks nicer.
The value of \verb|PhotonN| is added to the number of half-waves calculated by the program.
If you set it to $+1$ (or $-1$) the parity is reversed.
After the call to \verb|Zigzag| this variable is automatically reset to its default value 0.
If you need a non-zero \verb|PhotonN| for the next call to \verb|Zigzag|,
you have to set it again.\\
\begin{minipage}[t]{0.48\textwidth}
\small
\begin{verbatim}
begin translate 0.1 0.1
    amove 0 0
    @Zigzag 2 0
end translate
\end{verbatim}
\end{minipage}
\begin{minipage}[t]{0.48\textwidth}
\small
\begin{verbatim}
begin translate 3.1 0.1
    amove 0 0
    PhotonN = -1
    @Zigzag 2 0
end translate
\end{verbatim}
\end{minipage}\\
\includegraphics{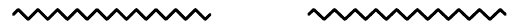}

\texttt{\bf @Photon \boldmath $x$ $y$}
draws a photon line from the current point to $(x,y)$.
It is controlled by \verb|PhotonL|, \verb|PhotonA|, \verb|PhotonN|,
just like zigzag lines.
\verb|PhotonN| is automatically reset to 0 after a call to \verb|Photon|.\\
\begin{minipage}[t]{0.48\textwidth}
\small
\begin{verbatim}
begin translate 0.1 0.6
    amove 0 0
    @Photon 2 0
end translate

begin translate 0.1 0.1
    amove 0 0
    PhotonA = -DefPhotonA
    @Photon 2 0
    PhotonA = DefPhotonA
end translate
\end{verbatim}
\end{minipage}
\begin{minipage}[t]{0.48\textwidth}
\small
\begin{verbatim}
begin translate 3.1 0.6
    amove 0 0
    PhotonL = 1.5*DefPhotonL
    @Photon 2 0
    PhotonL = DefPhotonL
end translate

begin translate 3.1 0.1
    amove 0 0
    PhotonA = 1.5*DefPhotonA
    @Photon 2 0
    PhotonA = DefPhotonA
end translate
\end{verbatim}
\end{minipage}\\
\includegraphics{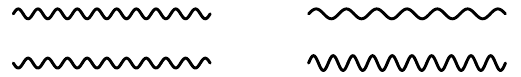}\\
\begin{minipage}[t]{0.48\textwidth}
\small
\begin{verbatim}
begin translate 0.1 0.1
    amove 0 0
    @Photon 2 0
end translate
\end{verbatim}
\end{minipage}
\begin{minipage}[t]{0.48\textwidth}
\small
\begin{verbatim}
begin translate 3.1 0.1
    amove 0 0
    PhotonN = -1
    @Photon 2 0
end translate
\end{verbatim}
\end{minipage}\\
\includegraphics{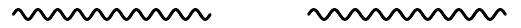}

\texttt{\bf @Gluon \boldmath $x$ $y$}
draws a gluon line from the current point to $(x,y)$.
Parametrically it is given by
\[
x = \frac{\lambda}{2\pi} \bigl[t + w (1 - \cos t)\bigr]\,,\quad
y = a \sin t\,.
\]
It is controlled by $\mathtt{PhotonL} = \lambda/2$, $\mathtt{PhotonA} = a$,
the winding coefficient $\mathtt{GluonW} = w$ (see the picture for its effect),
and \verb|GluonS|.
If $\mathtt{GluonS}>0$ (default),
both ends of the gluon line are of the same kind;
if $\mathtt{GluonS}<0$,
the beginning of the line looks as it should for given \verb|GluonW|, \verb|PhotonA|,
but its end looks as if the signs of these parameters has changed
(it may be useful for smooth joining two gluon lines).\\
\begin{minipage}[t]{0.48\textwidth}
\small
\begin{verbatim}
begin translate 0.1 1.1
    amove 0 0
    @Gluon 2 0
end translate

begin translate 2.6 1.1
    amove 0 0
    PhotonA = -DefPhotonA
    @Gluon 2 0
end translate

begin translate 5.1 1.1
    amove 0 0
    PhotonA = 1.5*DefPhotonA
    @Gluon 2 0
    PhotonA = DefPhotonA
end translate

begin translate 0.1 0.6
    amove 0 0
    PhotonL = 0.75*DefPhotonL
    @Gluon 2 0
    PhotonL = DefPhotonL
end translate
\end{verbatim}
\end{minipage}
\begin{minipage}[t]{0.48\textwidth}
\small
\begin{verbatim}
begin translate 2.6 0.6
    amove 0 0
    GluonW = 0.75*DefGluonW
    @Gluon 2 0
end translate

begin translate 5.1 0.6
    amove 0 0
    GluonW = 1.25*DefGluonW
    @Gluon 2 0
end translate

begin translate 0.1 0.1
    amove 0 0
    GluonW = -DefGluonW
    @Gluon 2 0
    GluonW = DefGluonW
end translate

begin translate 2.6 0.1
    amove 0 0
    GluonS = -1
    @Gluon 2 0
end translate

begin translate 5.1 0.1
    amove 0 0
    GluonW = -DefGluonW
    PhotonA = -DefPhotonA
    @Gluon 2 0
end translate
\end{verbatim}
\end{minipage}\\
\includegraphics{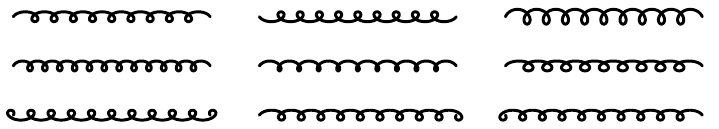}

\verb|Arrow|, \verb|Mom|, \verb|DMom| can be used with all kinds of lines
(including the invisible line \verb|Feyn|).
However, \verb|Arrow| does not look nice with \verb|Zigzag|, \verb|Photon|, \verb|Gluon|;
in these cases, it is better to use \verb|Mom| (or \verb|DMom|).

\texttt{\bf @Shadow \boldmath $f$}
constructs a rectangle of length $2*\mathtt{ShadowL}$ and width $2*\mathtt{ShadowA}$
whose center is the point at the fraction $f$ of the current line length.
If fills this rectangle with the color \verb|ShadowC$| and redraws the current line.
When two lines of a diagram intersect but there is no vertex
(e.\,g., in a non-planar diagram),
some authors prefer to make a small gap in one of these two lines.
\verb|Shadow| can be used to make such a gap
(the color \verb|ShadowC$| should coincide with the background color,
usually \verb|"white"|, therefore this is its default value;
sometimes in presentations people draw diagrams on non-white backgrounds,
then \verb|ShadowC$| should be set accordingly).
Here, for illustration, we set \verb|ShadowC$| to light-gray,
so that this rectangle becomes visible.\\
\begin{minipage}[t]{0.48\textwidth}
\small
\begin{verbatim}
begin translate 0.6 0.6
    amove 0 -0.5
    @Fermion 0 0.5
    amove -0.5 0
    @Fermion 0.5 0
    @Shadow 0.5
end translate

begin translate 2.6 0.6
    amove 0 -0.5
    @Fermion 0 0.5
    amove -0.5 0
    @Fermion 0.5 0
    ShadowC$ = "gray10"
    @Shadow 0.5
end translate
\end{verbatim}
\end{minipage}
\begin{minipage}[t]{0.48\textwidth}
\small
\begin{verbatim}
begin translate 4.6 0.6
    amove 0 -0.5
    @Fermion 0 0.5
    amove -0.5 0
    @Fermion 0.5 0
    ShadowL = 1.5*DefShadowL
    @Shadow 0.5
    ShadowL = DefShadowL
end translate

begin translate 6.6 0.6
    amove 0 -0.5
    @Fermion 0 0.5
    amove -0.5 0
    @Fermion 0.5 0
    ShadowA = 1.5*DefShadowA
    @Shadow 0.5
    ShadowA = DefShadowA
end translate
\end{verbatim}
\end{minipage}\\
\includegraphics{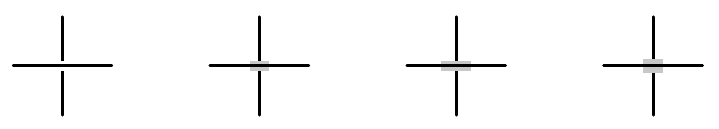}\\
If you mark all real interactions by \verb|Vert|
so that all intersections of lines without \verb|Vert| are spurious
then there is no need to use \verb|Shadow|.

\section{Arcs and circles}
\label{S:Arc}

\texttt{\bf @Fermion2 \boldmath $x_1$ $y_1$ $x$ $y$}
draws a fermion line --- an arc from the current point via $(x_1,y_1)$ to $(x,y)$.\\
\begin{minipage}[t]{0.48\textwidth}
\small
\begin{verbatim}
begin translate 0.1 0.7
    x1 = 1
    y1 = 0.75
    x = 2
    y = 0
    amove 0 0
    @Vert
    @Fermion2 x1 y1 x y
    @Vert
    @Arrow 1/4
    @Arrow 3/4
\end{verbatim}
\end{minipage}
\begin{minipage}[t]{0.48\textwidth}
\small
\begin{verbatim}
    amove x1 y1
    @Vert

    amove x1 y1+0.1
    set just bc
    tex "$(x_1,y_1)$"

    amove x y-0.1
    set just tc
    tex "$(x,y)$"
end translate
\end{verbatim}
\end{minipage}\\
\includegraphics{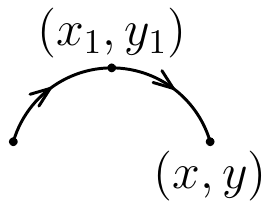}\\
\verb|Arrow|, \verb|Mom|, \verb|DMom|, \verb|Shadow|
can be used with arc-shaped lines.

\texttt{\bf @Feyn2 \boldmath $x_1$ $y_1$ $x$ $y$}
draws an invisible arc from the current point via $(x_1,y_1)$ to $(x,y)$.
It can be used to make \verb|Arrow|, \verb|Mom| or \verb|DMom|
point in the opposite direction.\\
\begin{minipage}[t]{0.48\textwidth}
\small
\begin{verbatim}
begin translate 0.1 0.1
    x1 = 1
    y1 = 0.75
    x = 2
    y = 0
    amove 0 0
    @Vert
    @Fermion2 x1 y1 x y
    @Vert
\end{verbatim}
\end{minipage}
\begin{minipage}[t]{0.48\textwidth}
\small
\begin{verbatim}
    @Arrow 1/4
    @Arrow 3/4
    @Feyn2 x1 y1 0 0
    MomD = -DefMomD
    @Mom 1/4
    @Mom 3/4
    MomD = DefMomD
    amove x1 y1
    @Vert
end translate
\end{verbatim}
\end{minipage}\\
\includegraphics{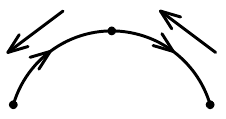}

\texttt{\bf @Double2 \boldmath $x_1$ $y_1$ $x$ $y$}
draws a double line from the current point via $(x_1,y_1)$ to $(x,y)$.\\
\begin{minipage}[t]{0.42\textwidth}
\small
\begin{verbatim}
x1 = 1
y1 = 0.75
x = 2
y = 0

begin translate 0.1 0.1
    amove 0 0
    @Double2 x1 y1 x y
end translate
\end{verbatim}
\end{minipage}
\begin{minipage}[t]{0.48\textwidth}
\small
\begin{verbatim}
begin translate 3.1 0.1
    amove 0 0
    c = (sqr(x1)-sqr(y1))/(2*x1*y1)
    DoubleB = c
    DoubleE = c
    @Double2 x1 y1 x y
end translate
\end{verbatim}
\end{minipage}\\
\includegraphics{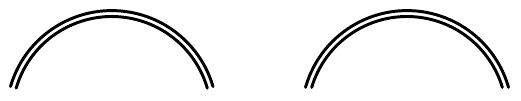}\\
In the right-hand figure,
the ends of the lines forming the double line are at the same height.

\texttt{\bf @Dash2 \boldmath $x_1$ $y_1$ $x$ $y$}
draws a dashed line from the current point via $(x_1,y_1)$ to $(x,y)$.\\
\begin{minipage}[t]{0.48\textwidth}
\small
\begin{verbatim}
x1 = 1
y1 = 0.75
x = 2
y = 0

begin translate 0.1 0.1
    amove 0 0
    @Vert
    @Dash2 x1 y1 x y
    @Vert
end translate

begin translate 3.1 0.1
    amove 0 0
    DashB = -1
    @Vert
    @Dash2 x1 y1 x y
    @Vert
    DashB = 1
end translate
\end{verbatim}
\end{minipage}
\begin{minipage}[t]{0.48\textwidth}
\small
\begin{verbatim}
begin translate 6.1 0.1
    amove 0 0
    DashE = -1
    @Vert
    @Dash2 x1 y1 x y
    @Vert
end translate

begin translate 9.1 0.1
    amove 0 0
    DashB = -1
    @Vert
    @Dash2 x1 y1 x y
    @Vert
    DashB = 1
    DashE = 1
end translate
\end{verbatim}
\end{minipage}\\
\includegraphics{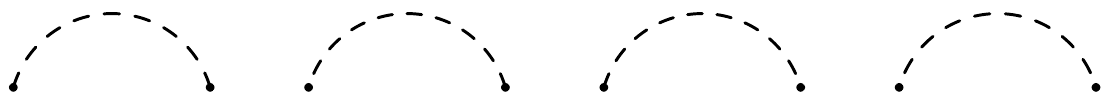}

\texttt{\bf @Dots2 \boldmath $x_1$ $y_1$ $x$ $y$}
draws a dotted line from the current point via $(x_1,y_1)$ to $(x,y)$.\\
\begin{minipage}[b]{0.48\textwidth}
\small
\begin{verbatim}
begin translate 0.1 0.1
    amove 0 0
    marker cross 0.5
    @Dots2 1 0.75 2 0
    marker cross 0.5
end translate
\end{verbatim}
\end{minipage}
\begin{minipage}[b]{0.48\textwidth}
\includegraphics{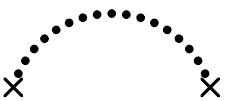}
\end{minipage}

\texttt{\bf @Zigzag2 \boldmath $x_1$ $y_1$ $x$ $y$}
draws a zigzag line from the current point via $(x_1,y_1)$ to $(x,y)$.\\
\begin{minipage}[t]{0.48\textwidth}
\small
\begin{verbatim}
begin translate 0.1 0.1
    amove 0 0
    @Zigzag2 1 0.75 2 0
end translate

begin translate 3.1 0.1
    amove 0 0
    PhotonA = -DefPhotonA
    @Zigzag2 1 0.75 2 0
    PhotonA = DefPhotonA
end translate
\end{verbatim}
\end{minipage}
\begin{minipage}[t]{0.48\textwidth}
\small
\begin{verbatim}
begin translate 6.1 0.1
    amove 0 0
    PhotonN = -1
    @Zigzag2 1 0.75 2 0
end translate
\end{verbatim}
\end{minipage}\\
\includegraphics{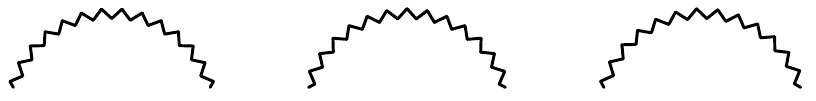}

\texttt{\bf @Photon2 \boldmath $x_1$ $y_1$ $x$ $y$}
draws a photon line from the current point via $(x_1,y_1)$ to $(x,y)$.\\
\begin{minipage}[t]{0.48\textwidth}
\small
\begin{verbatim}
begin translate 0.1 0.1
    amove 0 0
    @Photon2 1 0.75 2 0
end translate

begin translate 3.1 0.1
    amove 0 0
    PhotonA = -DefPhotonA
    @Photon2 1 0.75 2 0
    PhotonA = DefPhotonA
end translate
\end{verbatim}
\end{minipage}
\begin{minipage}[t]{0.48\textwidth}
\small
\begin{verbatim}
begin translate 6.1 0.1
    amove 0 0
    PhotonN = -1
    @Photon2 1 0.75 2 0
end translate
\end{verbatim}
\end{minipage}\\
\includegraphics{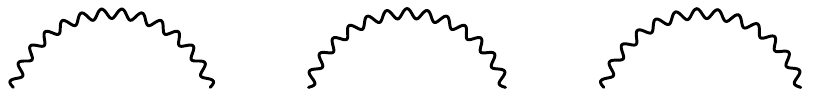}

\texttt{\bf @Gluon2 \boldmath $x_1$ $y_1$ $x$ $y$}
draws a gluon line from the current point via $(x_1,y_1)$ to $(x,y)$.\\
\begin{minipage}[t]{0.48\textwidth}
\small
\begin{verbatim}
begin translate 0.1 0.1
    amove 0 0
    @Gluon2 1 0.75 2 0
end translate
\end{verbatim}
\end{minipage}
\begin{minipage}[t]{0.48\textwidth}
\small
\begin{verbatim}
begin translate 3.1 0.1
    amove 0 0
    PhotonA = -DefPhotonA
    @Gluon2 1 0.75 2 0
    PhotonA = DefPhotonA
end translate
\end{verbatim}
\end{minipage}\\
\includegraphics{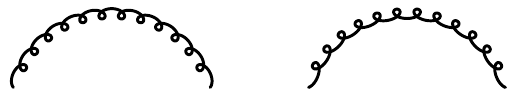}

\texttt{\bf @Mom2 \boldmath $f$}
draws an arc-shaped arrow for the momentum of an arc-shaped line
at the fraction $f$ of its length.
\texttt{\bf @DMom2 \boldmath $f$}
draws an arc-shaped double arrow.\\
\begin{minipage}[t]{0.48\textwidth}
\small
\begin{verbatim}
begin translate 0.1 0.1
    amove 0 0
    @fermion2 1 0.75 2 0
    @Mom2 1/4
    MomD = -DefMomD
    @Mom2 3/4
    MomD = DefMomD
end translate
\end{verbatim}
\end{minipage}
\begin{minipage}[t]{0.48\textwidth}
\small
\begin{verbatim}
begin translate 3.1 0.1
    amove 0 0
    @fermion2 1 0.75 2 0
    @DMom2 1/4
    MomD = -DefMomD
    @DMom2 3/4
    MomD = DefMomD
end translate
\end{verbatim}
\end{minipage}\\
\includegraphics{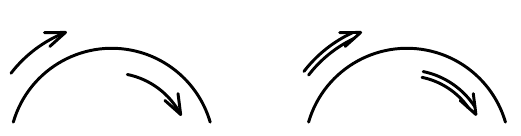}

\texttt{\bf @Shadow2 \boldmath $f$}
is similar to \verb|Shadow|, but the filled region
is bounded by arcs.\\
\begin{minipage}[t]{0.48\textwidth}
\small
\begin{verbatim}
begin translate 0.1 0.1
    amove 1 0
    @Fermion 1 1
    amove 0 0
    @Fermion2 1 0.75 2 0
    ShadowC$ = "gray10"
\end{verbatim}
\end{minipage}
\begin{minipage}[t]{0.48\textwidth}
\small
\begin{verbatim}
    ShadowL = 3*DefShadowL
    ShadowA = 2*DefShadowA
    @Shadow2 1/2
    ShadowC$ = "white"
    ShadowL = DefShadowL
    ShadowA = DefShadowA
end translate
\end{verbatim}
\end{minipage}\\
\includegraphics{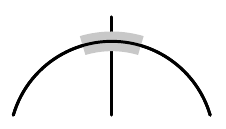}

\texttt{\bf @FermionC \boldmath $x$ $y$}
draws a fermion line --- circle with the center $(x,y)$ via the current point.
Its orientation is determined by \verb|CircleD|:
$+1$ means counterclockwise (the default), $-1$ means clockwise.
It is important for \verb|Arrow| and similar objects.
If you want, say, \verb|Arrow| and \verb|Mom| to point in opposite directions,
you can use \texttt{\bf @FeynC \boldmath $x$ $y$} --- an invisible circle.\\
\begin{minipage}[t]{0.48\textwidth}
\small
\begin{verbatim}
begin translate 1.1 1.1
    amove 0 -1
    @Vert
    @FermionC 0 0
    @Arrow 1/2
end translate

begin translate 4.1 1.1
    amove 0 -1
    @Vert
    CircleD = -1
    @FermionC 0 0
    CircleD = 1
    @Arrow 1/2
end translate
\end{verbatim}
\end{minipage}
\begin{minipage}[t]{0.48\textwidth}
\small
\begin{verbatim}
begin translate 7.1 1.1
    amove 0 -1
    @Vert
    CircleD = -1
    @FermionC 0 0
    CircleD = 1
    @Arrow 1/2
    @FeynC 0 0
    MomD = -DefMomD
    @Mom 1/2
    MomD = DefMomD
end translate
\end{verbatim}
\end{minipage}\\
\includegraphics{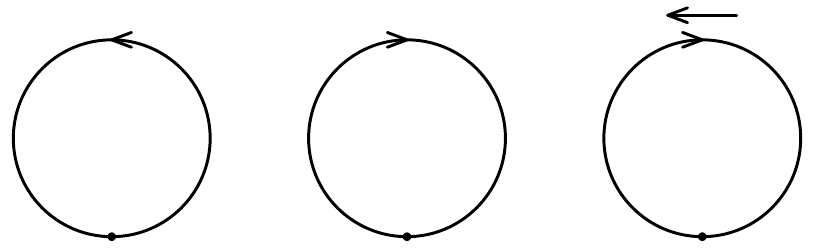}

\texttt{\bf @DoubleC \boldmath $x$ $y$}
draws a double line --- circle with the center $(x,y)$ via the current point.\\
\begin{minipage}[t]{0.48\textwidth}
\small
\begin{verbatim}
begin translate 1.1 1.1
    amove 0 -1
    @Vert
    @DoubleC 0 0
    @Arrow 1/2
end translate
\end{verbatim}
\end{minipage}
\begin{minipage}[t]{0.48\textwidth}
\small
\begin{verbatim}
begin translate 4.1 1.1
    amove 0 -1
    @Vert
    CircleD = -1
    @DoubleC 0 0
    CircleD = 1
    @Arrow 1/2
end translate
\end{verbatim}
\end{minipage}\\
\includegraphics{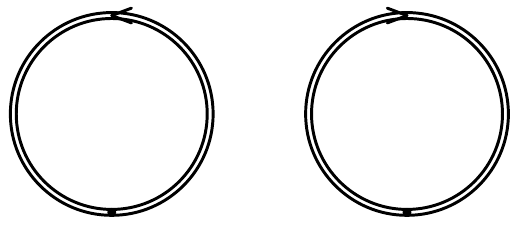}

\texttt{\bf @DashC \boldmath $x$ $y$}
draws a dashed line --- circle with the center $(x,y)$ via the current point.
With the default setting $\mathtt{DashB} = \mathtt{DashE} =1$
the dash around the beginning (or the end --- it's the same point) of the line
is 2 times longer than any other dash.
This is not very nice;
you can change \verb|DashB| and/or \verb|DashE| to obtain the result you want.\\
\begin{minipage}[t]{0.48\textwidth}
\small
\begin{verbatim}
begin translate 1.1 1.1
    amove 0 -1
    @Vert
    @DashC 0 0
end translate

begin translate 4.1 1.1
    amove 0 -1
    @Vert
    DashE = -1
    @DashC 0 0
end translate
\end{verbatim}
\end{minipage}
\begin{minipage}[t]{0.48\textwidth}
\small
\begin{verbatim}
begin translate 7.1 1.1
    amove 0 -1
    @Vert
    DashB = 0.5
    DashE = 0.5
    @DashC 0 0
    DashB = 1
    DashE = 1
end translate
\end{verbatim}
\end{minipage}\\
\includegraphics{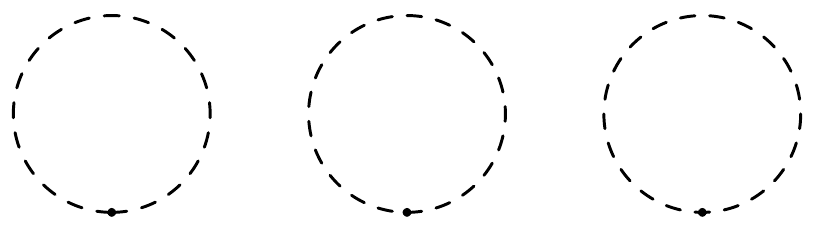}

\texttt{\bf @DotsC \boldmath $x$ $y$}
draws a dotted line --- circle with the center $(x,y)$ via the current point.
It does not draw the dot at the beginning (or end --- it's the same point) of the line;
if you want it, use \verb|@Vert|.\\
\begin{minipage}[b]{0.48\textwidth}
\small
\begin{verbatim}
begin translate 1.1 1.1
    amove 0 -1
    @Vert
    @DotsC 0 0
end translate
\end{verbatim}
\end{minipage}
\begin{minipage}[b]{0.48\textwidth}
\includegraphics{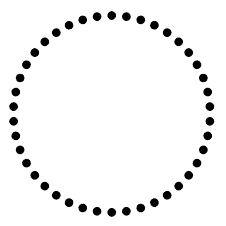}
\end{minipage}

\texttt{\bf @ZigzagC \boldmath $x$ $y$}
draws a zigzag line --- circle with the center $(x,y)$ via the current point.
If there is a half-integer number of waves in the circle, the result looks not very nice.
It can be improved by setting $\mathtt{PhotonN} = 1$ (or $-1$).\\
\begin{minipage}[t]{0.48\textwidth}
\small
\begin{verbatim}
begin translate 1.1 1.1
    amove 0 -1
    @ZigzagC 0 0
end translate
\end{verbatim}
\end{minipage}
\begin{minipage}[t]{0.48\textwidth}
\small
\begin{verbatim}
begin translate 4.1 1.1
    amove 0 -1
    PhotonN = -1
    @ZigzagC 0 0
end translate
\end{verbatim}
\end{minipage}\\
\includegraphics{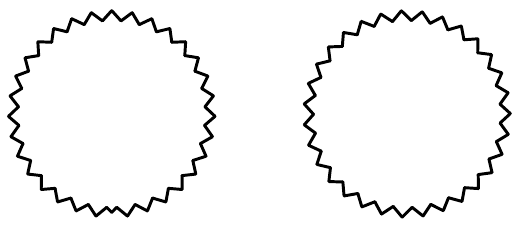}

\texttt{\bf @PhotonC \boldmath $x$ $y$}
draws a photon line --- circle with the center $(x,y)$ via the current point.
If there is a half-integer number of waves in the circle, the result looks not very nice.
It can be improved by setting $\mathtt{PhotonN} = 1$ (or $-1$).\\
\begin{minipage}[t]{0.48\textwidth}
\small
\begin{verbatim}
begin translate 1.1 1.1
    amove 0 -1
    @PhotonC 0 0
end translate
\end{verbatim}
\end{minipage}
\begin{minipage}[t]{0.48\textwidth}
\small
\begin{verbatim}
begin translate 4.1 1.1
    amove 0 -1
    PhotonN = -1
    @PhotonC 0 0
end translate
\end{verbatim}
\end{minipage}\\
\includegraphics{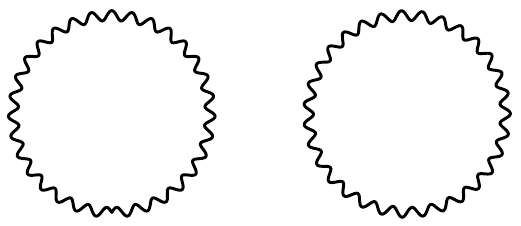}

\texttt{\bf @GluonC \boldmath $x$ $y$}
draws a gluon line --- circle with the center $(x,y)$ via the current point.
For a nice look, it is useful to set $\mathtt{GluonS} = -1$.\\
\begin{minipage}[t]{0.48\textwidth}
\small
\begin{verbatim}
begin translate 1.1 1.1
    amove 0 -1
    GluonS = -1
    @GluonC 0 0
end translate
\end{verbatim}
\end{minipage}
\begin{minipage}[t]{0.48\textwidth}
\small
\begin{verbatim}
begin translate 4.1 1.1
    amove 0 -1
    PhotonA = -DefPhotonA
    @GluonC 0 0
    PhotonA = DefPhotonA
    GluonS = 1
end translate
\end{verbatim}
\end{minipage}\\
\includegraphics{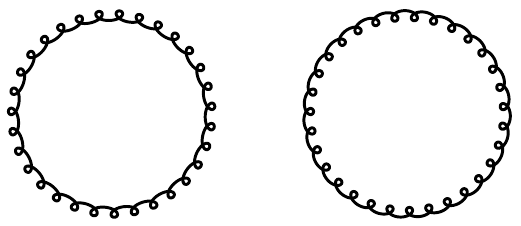}

If the current point, $(x_1,y_1)$, and $(x,y)$ lie on a straight line,
\texttt{\bf @Fermion2 \boldmath $x_1$ $y_1$ $x$ $y$}
works as
\texttt{\bf @Fermion \boldmath $x$ $y$}.\\
\begin{minipage}[b]{0.48\textwidth}
\small
\begin{verbatim}
begin translate 0.1 0.1
    amove 0 0
    @Vert
    @Fermion2 1 0 2 0
    @Vert
    amove 1 0
    @Vert
end translate
\end{verbatim}
\end{minipage}
\begin{minipage}[b]{0.48\textwidth}
\includegraphics{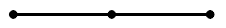}
\end{minipage}\\
If the point $(x,y)$ coincides with the current point,
\texttt{\bf @Fermion2 \boldmath $x_1$ $y_1$ $x$ $y$}
works as
\texttt{\bf @FermionC \boldmath $x'$ $y'$},
where the center $(x',y')$ is in the middle between the current point and $(x_1,y_1)$.
In this (and only this) case \verb|CircleD| influences its direction.\\
\begin{minipage}[t]{0.48\textwidth}
\small
\begin{verbatim}
begin translate 0.1 1.1
    amove 0 0
    @Fermion2 2 0 0 0
end translate
\end{verbatim}
\end{minipage}
\begin{minipage}[t]{0.48\textwidth}
\small
\begin{verbatim}
begin translate 3.1 1.1
    amove 0 0
    CircleD = -1
    @Fermion2 2 0 0 0
    CircleD = 1
    @Arrow 1/2
end translate
\end{verbatim}
\end{minipage}\\
\includegraphics{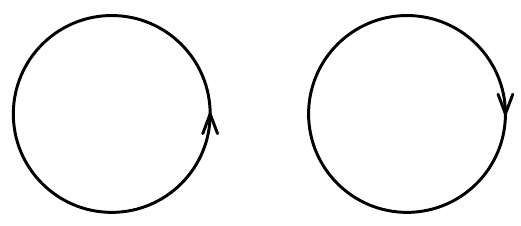}\\
The same is true for \verb|Feyn2|, \verb|Double2|, \verb|Dash2|, \verb|Dots2|, \verb|Zigzag2|, \verb|Photon2|, \verb|Gluon2|.

\section{Conclusion}
\label{S:Conc}

Here is the summary table of all objects and control parameters:
$+$ means that this parameter influences this object,
$\pm$ --- only in rare special cases (see the main text).
Line width can be updated by the builtin command \verb|set lwidth|.
For a parameter \verb|Param|,
$+$ in the row \textit{Def} means that the variable \verb|DefParam| is defined;
$+$ in the row \textit{Autoreset} means that this parameter is automatically reset
to its default value after a call to the procedure which uses this parameter.\\
\scriptsize
\begin{tabular}{|l|c|c|c|c|c|c|c|c|c|c|c|c|c|c|c|c|c|c|c|}
\hline
&
\rotatebox{90}{\texttt{lwidth}} &
\rotatebox{90}{\texttt{DoubleA}} &
\rotatebox{90}{\texttt{DoubleB}} \rotatebox{90}{\texttt{DoubleE}} &
\rotatebox{90}{\texttt{DashL}} \rotatebox{90}{\texttt{DashF}} &
\rotatebox{90}{\texttt{DashB}} \rotatebox{90}{\texttt{DashE}} &
\rotatebox{90}{\texttt{DashN}} &
\rotatebox{90}{\texttt{DotR}} &
\rotatebox{90}{\texttt{DotsL}} &
\rotatebox{90}{\texttt{FillC\$}} &
\rotatebox{90}{\texttt{PhotonL}} \rotatebox{90}{\texttt{PhotonA}} &
\rotatebox{90}{\texttt{PhotonN}} &
\rotatebox{90}{\texttt{GluonW}} &
\rotatebox{90}{\texttt{GluonS}} &
\rotatebox{90}{\texttt{ArrowL}} \rotatebox{90}{\texttt{ArrowA}} &
\rotatebox{90}{\texttt{ArrowD}} &
\rotatebox{90}{\texttt{MomL}} \rotatebox{90}{\texttt{MomD}} &
\rotatebox{90}{\texttt{ShadowL}} \rotatebox{90}{\texttt{ShadowA}} &
\rotatebox{90}{\texttt{ShadowC\$}} &
\rotatebox{90}{\texttt{CircleD}}\\
\hline
\verb|Feyn|  &&&&&&&&&&&&&&&&&&&\\
\verb|Feyn2| &&&&&&&&&&&&&&&&&&& $\pm$ \\
\verb|FeynC| &&&&&&&&&&&&&&&&&&& $+$ \\
\hline
\verb|Fermion|  & $+$ &&&&&&&&&&&&&&&&&&\\
\verb|Fermion2| & $+$ &&&&&&&&&&&&&&&&&& $\pm$ \\
\verb|FermionC| & $+$ &&&&&&&&&&&&&&&&&& $+$ \\
\hline
\verb|Double|  & $+$ & $+$ & $+$ &&&&&&&&&&&&&&&&\\
\verb|Double2| & $+$ & $+$ & $+$ &&&&&&&&&&&&&&&& $\pm$ \\
\verb|DoubleC| & $+$ & $+$ &&&&&&&&&&&&&&&&& $+$ \\
\hline
\verb|Dash|  & $+$ &&& $+$ & $+$ & $+$ &&&&&&&&&&&&&\\
\verb|Dash2| & $+$ &&& $+$ & $+$ & $+$ &&&&&&&&&&&&& $\pm$ \\
\verb|DashC| & $+$ &&& $+$ & $+$ & $+$ &&&&&&&&&&&&& $+$ \\
\hline
\verb|Dots|  & $+$ &&&&&& $+$ & $+$ & $+$ &&&&&&&&&&\\
\verb|Dots2| & $+$ &&&&&& $+$ & $+$ & $+$ &&&&&&&&&& $\pm$ \\
\verb|DotsC| & $+$ &&&&&& $+$ & $+$ & $+$ &&&&&&&&&& $+$ \\
\hline
\verb|Zigzag|  & $+$ &&&&&&&&& $+$ & $+$ &&&&&&&&\\
\verb|Zigzag2| & $+$ &&&&&&&&& $+$ & $+$ &&&&&&&& $\pm$ \\
\verb|ZigzagC| & $+$ &&&&&&&&& $+$ & $+$ &&&&&&&& $+$ \\
\hline
\verb|Photon|  & $+$ &&&&&&&&& $+$ & $+$ &&&&&&&&\\
\verb|Photon2| & $+$ &&&&&&&&& $+$ & $+$ &&&&&&&& $\pm$ \\
\verb|PhotonC| & $+$ &&&&&&&&& $+$ & $+$ &&&&&&&& $+$ \\
\hline
\verb|Gluon|  & $+$ &&&&&&&&& $+$ && $+$ & $+$ &&&&&&\\
\verb|Gluon2| & $+$ &&&&&&&&& $+$ && $+$ & $+$ &&&&&& $\pm$ \\
\verb|GluonC| & $+$ &&&&&&&&& $+$ && $+$ & $+$ &&&&&& $+$ \\
\hline
\verb|Vert|    & $+$ &&&&&& $+$ && $+$ &&&&&&&&&&\\
\verb|Arrow|   & $+$ &&&&&&&&&&&&& $+$ & $+$ &&&&\\
\verb|Mom|     & $+$ &&&&&&&&&&&&& $+$ && $+$ &&&\\
\verb|Mom2|    & $+$ &&&&&&&&&&&&& $+$ && $+$ &&&\\
\verb|DMom|    & $+$ & $+$ &&&&&&&&&&&& $+$ && $+$ &&&\\
\verb|DMom2|   & $+$ & $+$ &&&&&&&&&&&& $+$ && $+$ &&&\\
\verb|Shadow|  &&&&&&&&&&&&&&&&& $+$ & $+$ &\\
\verb|Shadow2| &&&&&&&&&&&&&&&&& $+$ & $+$ &\\
\hline
\textit{Def}       & $+$ & $+$ && $+$ &&& $+$ & $+$ && $+$ && $+$ && $+$ && $+$ & $+$ &&\\
\textit{Autoreset} &&& $+$ &&& $+$ &&&&& $+$ &&&&&&&&\\
\hline
\end{tabular}
\normalsize

GLE has many tools for producing graphics,
and they may be combined with those of \verb|feyn.gle|.
Some special kinds of vertices can be denoted by markers,
as demonstrated in the example with \verb|@Dots|.
If you need more flexibility, you can define a subroutine for your vertices.
You can use standard GLE tools to produce shaded blobs of various shapes.\\
\begin{minipage}[t]{0.48\textwidth}
\small
\begin{verbatim}
x = 1.5
x1 = 1.8
r = 0.4
d = 0.1

size 2*(x1+d) x+r+2*d
include feyn.gle

begin translate x1+d x+r+d
    amove -x1 -x1
    @double 0 0
    @Vert
    @double x1 -x1
    amove 0 -x
    circle r fill shade1
\end{verbatim}
\end{minipage}
\begin{minipage}[t]{0.48\textwidth}
\small
\begin{verbatim}
    c = sqrt(0.5)*r
    amove x -x
    @Vert
    @Photon r -x
    amove 0.5*x -0.5*x
    @Vert
    @Photon c c-x
    amove -x -x
    @Vert
    PhotonA = -DefPhotonA
    @Photon -r -x
    amove -0.5*x -0.5*x
    @Vert
    @Photon -c c-x
end translate
\end{verbatim}
\end{minipage}\\
\includegraphics{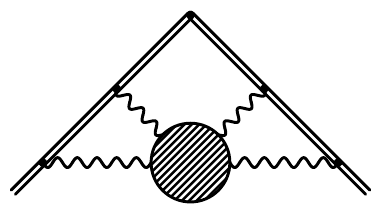}\\
GLE is a full-fledged programming language with subroutines, loops,
conditional statements, and all its facilities can be used in programs
for producing Feynman diagrams.
Of course, in order to use them one has to read the GLE manual
in addition to the current article.

The author has used this package for drawing many hundreds Feynman diagrams,
including a large number of articles, 3 books, many pdf presentations for conference talks,
and several lecture courses for students.
GLE with \verb|feyn.gle| works quite efficiently:
even for a complicated diagram the run time is typically $< 1\,\mathrm{sec}$.

All \verb|.gle| files used for pictures in this manual are attached to the arXiv preprint.

The work was supported by the Russian Ministry of Science and Higher Education.

\bibliographystyle{elsarticle-num}
\bibliography{feyn}

\end{document}